\documentclass[preprint]{jpsj3}
\usepackage{txfonts}

\title{Direct Observation of the Quantum Phase Transition of SrCu$_{2}$(BO$_{3}$)$_{2}$ by High-Pressure and Terahertz Electron Spin Resonance}

\author{Takahiro Sakurai$^1$\thanks{tsakurai@kobe-u.ac.jp},
Yuki Hirao$^2$,
Keigo Hijii$^3$,
Susumu Okubo$^3$,
Hitoshi Ohta$^3$,
Yoshiya Uwatoko$^4$,
Kazutaka Kudo$^5$,
and Yoji Koike$^6$}
\inst{$^1$Research Facility Center for Science and Technology, Kobe University, Kobe 657-8501, Japan\\
$^2$Graduate School of Science, Kobe University, Kobe 657-8501, Japan\\
$^3$Molecular Photoscience Research Center, Kobe University, Kobe 657-8051, Japan\\
$^4$Institute for Solid State Physics, University of Tokyo, Chiba 277-8581, Japan\\
$^5$Research Institute for Interdisciplinary Science, Okayama University, Okayama 700-8530, Japan\\
$^6$Department of Applied Physics, Tohoku University, Sendai 980-8579, Japan} %\\

\abst{High-pressure and high-field electron spin resonance (ESR) measurements have been performed on a single crystal of the orthogonal-dimer spin system SrCu$_{2}$(BO$_{3}$)$_{2}$.
With frequencies below 1 THz, ESR signals associated with transitions from the singlet ground state to the one-triplet excited states and the two-triplet bound state were observed at pressures up to 2.1 GPa.
We obtained directly the pressure dependence of the gap energies, finding a clear first-order phase transition at $P_{c}=1.85\pm0.05$ GPa.
By comparing this pressure dependence with the calculated excitation energies obtained from an exact diagonalization, we determined the precise pressure dependence for inter- ($J'$) and intra-dimer  ($J$)  exchange interactions considering the Dzyaloshinski-Moriya interaction.
Thus this system undergoes a first-order quantum phase transition from the dimer singlet phase to a plaquette singlet phase above the ratio $(J'/J)_{c}=0.660\pm0.003$.}

\kword{quantum phase transition, frustration, electron spin resonance, pressure, SrCu$_{2}$(BO$_{3}$)$_{2}$}
\begin{document}
\maketitle

Understanding the ground states of frustrated systems that have a large number of nearly degenerated states is one of the central issues in physics.
The importance of frustration has been recognized in a variety of condensed matter recently, such as magnetic materials \cite{frustration1, frustration2}, superconductors \cite{anderson, super} and ferroelectric materials \cite{ferroele1,ferroele2}.
Quantum magnets are representative materials exhibiting exotic ground states that arise from competition between spin frustration and quantum fluctuation \cite{frustration2}.
Among them, the orthogonal-dimer spin system of strontium copper borate SrCu$_{2}$(BO$_{3}$)$_{2}$ stands out markedly from other frustration systems because of its unique spin arrangement, known as the Shastry-Sutherland lattice \cite{kageyama, miyahara}.
The $S = 1/2$ antiferromagnetic dimers arrange orthogonally in a two-dimensional plane and also couple antiferromagnetically.
The Hamiltonian of this system is expressed as
\begin{equation}
\label{eq1}
{\cal H}_{0}=J\sum_{nn}{\bf S}_{i}\cdot{\bf S}_{j}+J'\sum_{nnn}{\bf S}_{i}\cdot{\bf S}_{j}
\end{equation}
where $nn$ and $nnn$ signify nearest-neighbor and next-nearest-neighbor, respectively.
The dimer singlet state is an exact eigenstate of this Hamiltonian (\ref{eq1}) \cite{SS}.
This is obviously the ground state under the large limit of the intradimer exchange interaction $J$, while the Hamiltonian (\ref{eq1}) with only interdimer exchange interaction $J'$ is equivalent to the two-dimensional square lattice for which the ground state is the N\'eel state.
There has been a long debate on an intermediate phase between these states because strong frustration within this system prevents theoretical analyses that approach the quantum critical point (QCP) precisely \cite{review, miyahara, koga, takushima, corboz}.

$J$ and the ratio $\alpha = J'/J$ are estimated to be $J=71-85$ K and $\alpha=0.60-0.64$ \cite {knetter, jaime, matsuda, miya2} for SrCu$_{2}$(BO$_{3}$)$_{2}$.
Through intensive studies on the intermediate phase, there is a consensus that a QCP of an intermediate phase exists around $\alpha_{c}(\rm theory)\sim0.68$ just above the  ratio of this compound and it is the plaquette singlet phase \cite{review, miyahara, koga, takushima, corboz}.
However, achieving experimentally the quantum phase transition (QPT) to the plaquette singlet phase has still been challenging. 

Pressure is the only way to explore the QPT of this compound.
Measurements of the magnetic susceptibility \cite{chi-T, INS}, NMR \cite{NMR, NMR2}, X-ray diffraction \cite{X-ray}, ESR \cite{ref12, ohta}, and inelastic neutron scattering (INS) \cite{INS} were performed under various pressures.
Temperature-dependent NMR measurements at 2.4 GPa indicate a change in symmetry from tetragonal to orthorhombic with decreasing temperature and a magnetic phase transition at 4 K \cite{NMR, NMR2}.
The spatially ordered state of two kinds of dimers with different spin gaps was proposed for this magnetic phase.
In contrast, from X-ray diffraction measurements for pressures up to 8 GPa, a collapse of the spin gap and the simultaneous second-order phase transition to the plaquette state at 2 GPa was proposed \cite{X-ray}.
However, the magnetic susceptibility \cite{chi-T} and ESR \cite{ref12, ohta} measurements below 2 GPa suggest that the spin gap remains open around 2 GPa.
A quite recent INS measurement also suggested that a QPT occurs between 1.60 and 2.15 GPa with remaining gap \cite{INS}.
Although, from experiment, the pressure-induced phase transition is most likely to exist around 2 GPa \cite{NMR, X-ray}, the behavior of its phase transition and the high pressure phase are still highly controversial.
The most difficult point in these experimental studies is that the quantitative change in the exchange interaction with pressure is unclear.
Zayed {\it et al}. \cite{INS} attempted an estimation of the pressure dependence of $\alpha$ by fitting the calculated result to the magnetic susceptibility data below 1 GPa.
However, since the large pressure drop in the clamped-type pressure cell from room to liquid helium temperature \cite{tompson} was not taken into account in this estimation, the obtained  $\alpha$ involves considerable error.
Therefore, the results cannot be compared directly with those from theory with varying $\alpha$.

In the present work, we have investigated the pressure effects on SrCu$_{2}$(BO$_{3}$)$_{2}$ from high-pressure ESR measurements.
We succeeded in observing the excitation gaps directly and in obtaining precise parameter values for $\alpha$ and $J$ under various pressures that are completely free from the pressure drop effect because the values were estimated at constant temperature. 
The system exhibits a clear phase transition at $P_{c}=1.85\pm0.05$ GPa, corresponding to $\alpha_{c} = 0.660\pm0.003$.
The conclusion is that in a comparison with theory \cite{takushima} the high-pressure phase is the plaquette phase.

\begin{figure}[t]
\includegraphics[width=1\linewidth]{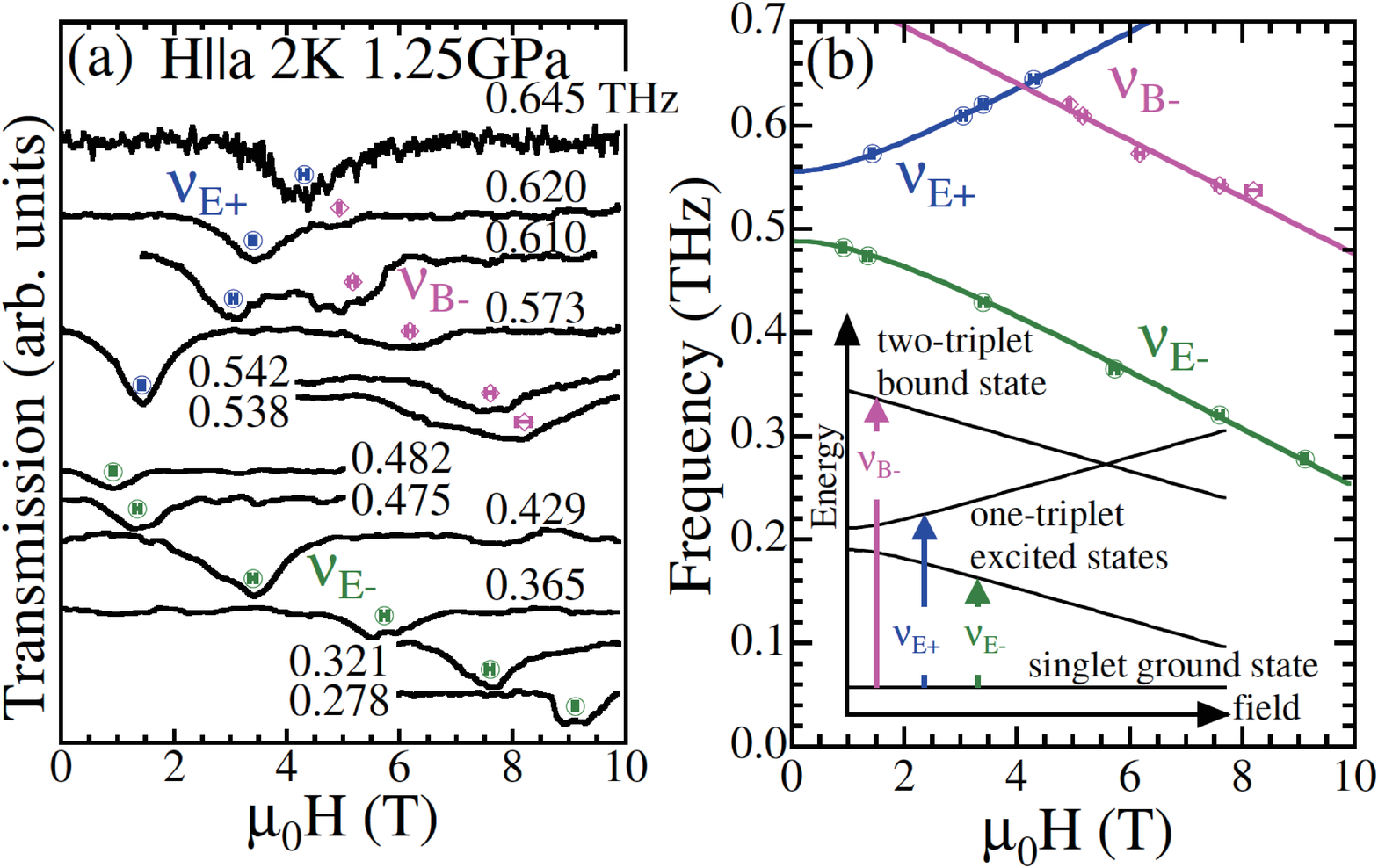}
\caption{\label{nama}(Color online) Typical frequency dependence of ESR spectra (a) and frequency-field diagram (b) obtained at 1.25 GPa and 2 K for H$\parallel$a. Inset:  schematic of the energy-field diagram and the observed ESR transitions.}
\end{figure} 
The terahertz ESR measurements at high pressure and high magnetic field have been performed by our recently developed system \cite{hybrid}.
A hybrid-type piston-cylinder pressure cell was used.
In this study, the frequency region extends from 0.08 to 0.80 THz.
Gunn oscillators and backward travelling wave oscillators were used as light sources.
The maximum pressure is 2.13 GPa.
The pressure is calibrated from relationship between the load at room temperature and the pressure around 3 K;
the accuracy is 0.02 GPa \cite{hybrid}.
A single crystal of SrCu$_{2}$(BO$_{3}$)$_{2}$ was grown by the traveling floating zone method.
The crystal axes were confirmed from X-ray diffraction measurements;
a sample of  dimensions 2$\times$2$\times$7 mm$^{3}$ cut along the $a$ axis was used.
The magnetic field was applied parallel to the $a$ axis.

Figure \ref{nama} (a) shows the typical frequency dependence of the ESR spectra with the ESR signals clearly visible (indicated by symbols).
The resonance fields are summarized in a frequency-field diagram [Fig. \ref{nama} (b)].
From the ESR measurements at ambient pressure and wide ranges of frequency and field \cite{nojiri}, the one-triplet excited states were found to lie above the singlet ground state and the triplet states of the two-triplet bound states lie further above these states [see inset of Fig. \ref{nama} (b)]. 
In our measurements, upper (+) and lower (-) ESR modes $\nu_{{\rm E}\pm}$ associated with the transition from the ground state to the one-triplet excited states and a lower ESR mode $\nu_{{\rm B}-}$ corresponding to the transition from the ground state to the lowest branch of triplet states of two-triplet bound states were observed.
For the ESR modes $\nu_{{\rm E}\pm}$, they have small splittings at zero field.
This zero-field splitting can be explained by the main component of the Dzyaloshinski-Moriya (DM) interaction of the system, that is, the $c$ component of the interdimer DM interaction.
The modes can also be well fitted by $h\nu_{{\rm E}\pm}=j\pm\sqrt{d^{2}+\left(g\mu_{\rm B}H\right)^{2}}$ \cite{cepas}, where $h$, $\nu$, $g$, $\mu_{\rm B}$, and $H$ are Planck's constant, frequency, $g$-value, Bohr magneton, and magnetic field, respectively, and $j$ and $d$ are fitting parameters.
From Fig. \ref{nama} (b), the $\nu_{{\rm E}\pm}$ modes are very well fitted with the $g$-value fixed at $g = 2.05$ \cite{nojiri}.
Moreover, the $\nu_{{\rm B}-}$ mode is fitted by a straight line with $g = 2.05$ \cite{nojiri}.
 
\begin{figure}[t]
\includegraphics[width=1\linewidth]{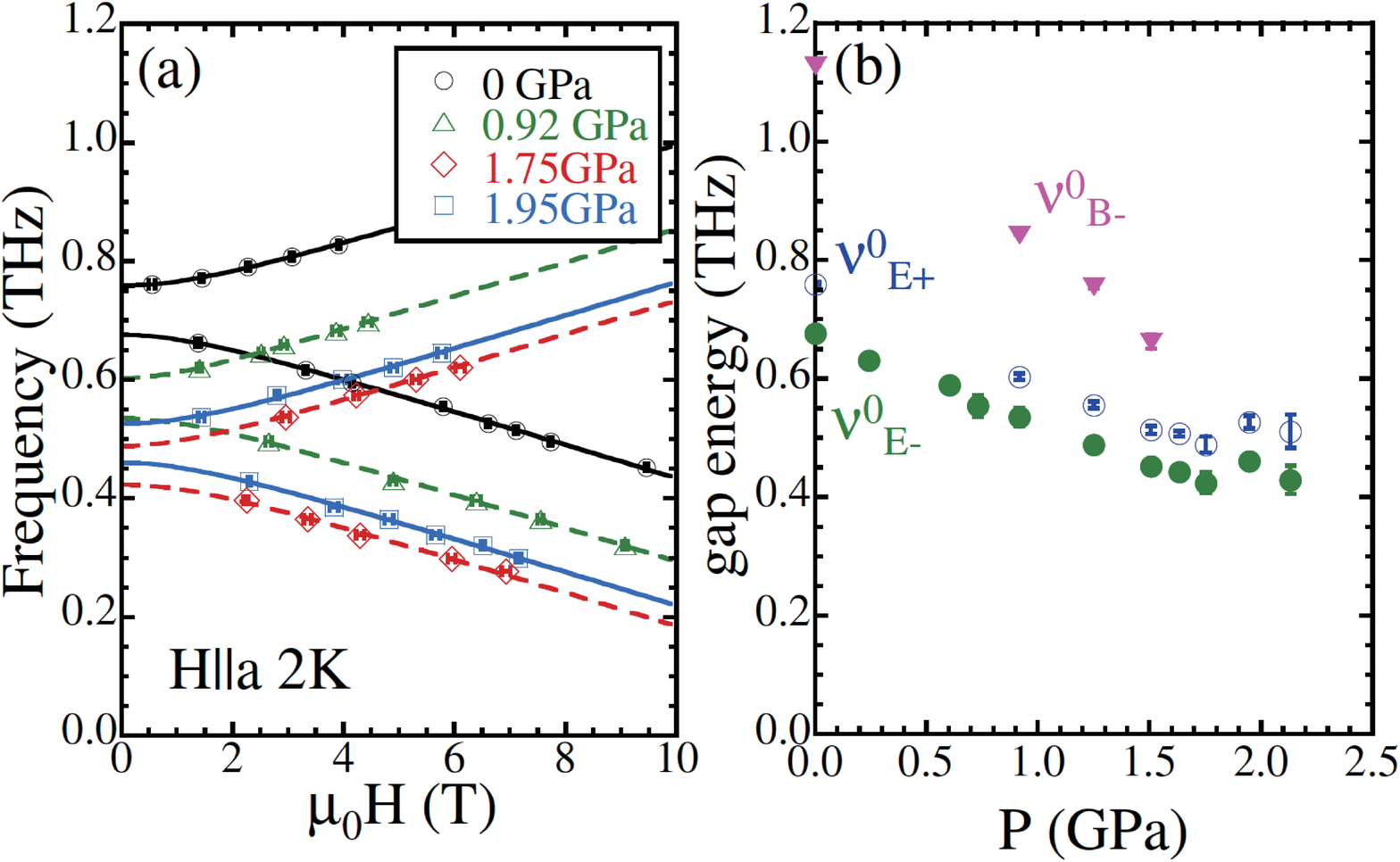}
\caption{\label{FH}(Color online) (a) Frequency-field diagram of the $\nu_{{\rm E}\pm}$ mode for various pressures. (b) Pressure dependence of the gap energies. The value $\nu_{{\rm B}-}^{0}$ at ambient pressure is taken from Ref. \cite{nojiri}.}
\end{figure}  
In Fig. \ref{FH} (a), the ESR modes $\nu_{{\rm E}\pm}$ obtained for various pressures are shown;
they are all well fitted by the same formula for the zero-field splitting.
In Fig. \ref{FH} (a), the gap energy at zero field is reduced with applied pressures below 1.75 GPa but begins to increase at 1.95 GPa.
Figure \ref{FH} (b) presents the pressure dependence of the gap energies at zero field, $\nu_{{\rm E}\pm}^{0}$ ($= j\pm d$) and $\nu_{{\rm B}-}^{0}$.
With increasing pressure, a reduction in the gap energy is seen with a discontinuous jump at 1.95 GPa.
Thus, we determined the transition pressure as $P_{c} = 1.85\pm0.05$ GPa.
Here we can exclude the possibility that this phase transition arises from a structural phase transition because the X-ray diffraction measurements obtained with applied pressures did not show any sign suggestive of a structural phase transition around 2 GPa \cite{X-ray, X-ray2}.
All absorption lines tend to be broadened around $P_{c}$ as pressure increases.
This is why the ESR mode $\nu_{{\rm B}-}$ could not be identified above 1.51 GPa  as its intensity is weaker than those of modes $\nu_{{\rm E}\pm}$ [Fig. \ref{nama} (a)], although it is not clear yet that this broadening comes from the intrinsic nature in this transition or the pressure inhomogeneous distribution.

We next extracted the exchange interactions from the obtained excitation energies \cite{suppl} to clarify the origin of this phase transition and the high-pressure phase.
The observed excitations are almost governed by the Hamiltonian in Eq. (\ref{eq1}).
However, to explain the fine splitting at zero field for the one-triplet excitation the DM interaction along the $c$ axis [${\cal H}_{\rm DM}=\sum_{nnn}{\bf D}_{ij}\cdot\left({\bf S}_{i}\times{\bf S}_{j}\right)$, ${\bf D}_{ij}=(0,0,\pm D)$] is required \cite{cepas}.
Therefore, we calculated the excitation energies for Hamiltonian ${\cal H}={\cal H}_{0}+{\cal H}_{\rm DM}$ including the three unknown parameters $J$, $J'$, and $D$ for 20 sites subject to periodic conditions by performing an exact diagonalization, and we compared the experimentally obtained excitation energies with those calculated.
For simplicity, the DM interaction is assumed to scale with the interdimer interaction ($D=kJ'$, $k$ is a parameter).

\begin{figure}[t]
\includegraphics[width=1\linewidth]{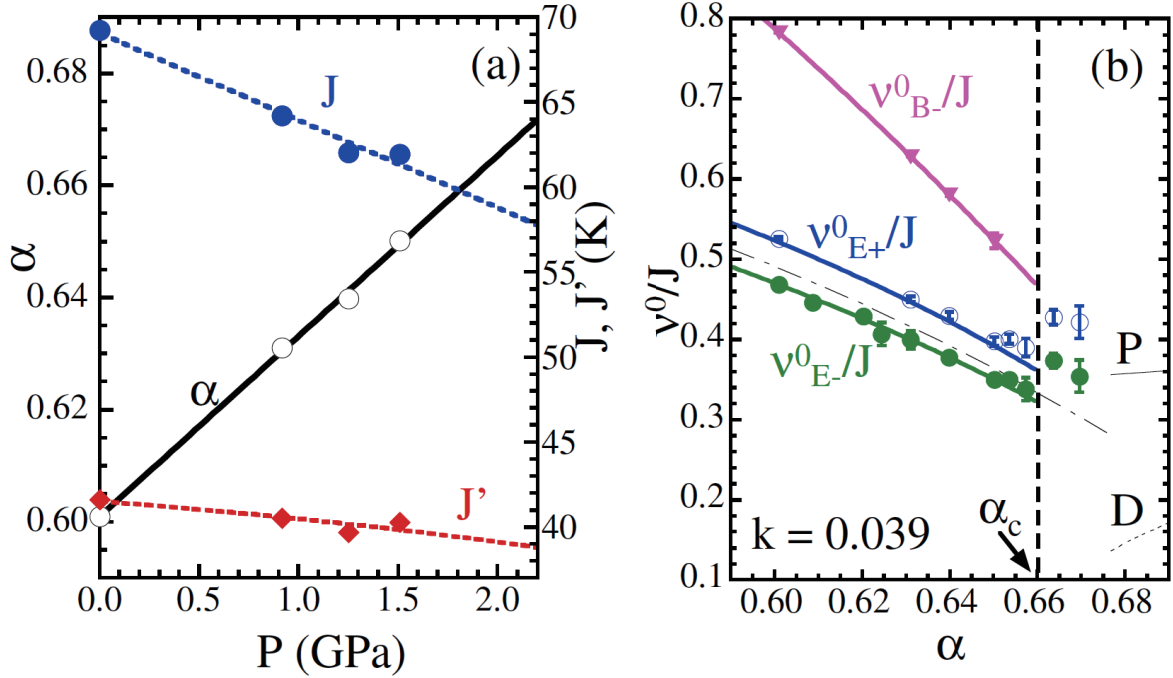}
\caption{\label{aJJ}(Color online)
(a) Pressure dependence of $J$, $J'$ and $\alpha$.
(b) Gap energies $\nu_{{\rm E}\pm}^{0}$ and $\nu_{{\rm B}-}^{0}$ normalized by $J$ as a function of $\alpha$. The symbols correspond to those in Fig. \ref{FH}(b). Thick lines are obtained by the exact diagonal calculation with $k=0.039$. Thin lines are the results obtained in Takushima {\it et al}. \cite{takushima}. See text for details.}
\end{figure}
At $P =$ 0, 0.92, 1.25, and 1.51 GPa, having obtained three excitation energies $\nu_{{\rm E}\pm}^{0}$ and $\nu_{{\rm B}-}^{0}$ [Fig. \ref{FH} (b)], we can uniquely determine the three unknown parameters $\alpha$, $J$, and $k$ without ambiguity.
Parameter $k$ was determined to be $k = 0.039$;
the obtained pressure dependence of $\alpha$ and $J$ is shown in Fig. \ref{aJJ} (a).
They are well fitted by straight lines giving expressions $J(P)/k_{\rm B}$[K] $= -5.14$[K/GPa]$P$[GPa]$+69.1$[K] and $\alpha(P)=0.0322P[{\rm GPa}]+0.601$.
With $J(P)$ and $\alpha(P)$ are fixed, we can estimate $J$ and $\alpha$ for other excitation data obtained at pressures except for the four pressures mentioned above.
We normalized the excitation energies by $J$ and plotted them as a function of $\alpha$ [Fig. \ref{aJJ} (b)].
Within the errors of uncertainty, the data show good agreement with the calculated results (thick lines) except for $\alpha>0.66$.
From the jump discontinuity, the transition point is obtained as $\alpha_{c}=0.660\pm0.003$.
Thus, we have determined precisely the pressure dependence of $J$, $J'$ ($=\alpha J$), and $D$ ($=kJ'$), and the transition point $\alpha_{c}$.
To the best of our knowledge, this is the first study to obtain such trends for the parameters of SrCu$_{2}$(BO$_{3}$)$_{2}$.

At ambient pressure, we obtained $J=69.1$ K, $\alpha=0.601$, and $D=1.6$ K, which agree well with the results obtained by comparing the excitation energies similarly from the INS measurement with those calculated using the perturbative unitary transformation ($J=71.5$ K, $\alpha=0.603$) \cite{knetter}.
The slight differences stem from the DM interaction that was taken into account in our calculation.
Although our estimation gives the smallest values among the estimations of $J$ and $\alpha$ for this compound \cite {knetter, jaime, matsuda, miya2}, they are the unique values that reproduced completely both energy gaps to the one-triplet excited states and the two-triplet bound state.
Moreover, our estimation explains the fine splitting of the one-triplet excitation at zero field.
Our parameter values are considered to be the most reliable.

We now compare our results with the theory by Takushima {\it et al}., who derived the triplet excitation energies for the plaquette phase \cite{takushima}.
Their results are marked as thin lines in Fig. \ref{aJJ} (b).
They calculated the excitation energy by series expansion using the fact that the plaquette phase of the Shastry-Sutherland lattice can be connected adiabatically to that in the 1/5-depleted square lattice and in the orthogonal dimer chain.
They obtained two triplet excitations:
one is an excitation that simply breaks the plaquette singlet [P-mode, thin solid line in Fig. \ref{aJJ} (b)] and the another is an excitation that has intermediate properties between the dimer singlet and the plaquette singlet (D-mode, thin dotted line).
Below $\alpha_{c}$, although the one-triplet excitation does not split as the DM interaction is not taken into account in their calculation, the gap energies coincide very well (thin broken line).
Above $\alpha_{c}$, the observed gap energies are consistent quantitatively with that of the P-mode in terms of the magnitude and the dependence on $\alpha$, although the transition point is slightly smaller than the theoretically obtained $\alpha_{c}({\rm theory})\sim0.68$ \cite{review, miyahara, koga, takushima, corboz}.
In particular, the magnitude of the jump of about $0.06\nu^{0}_{E\pm}/J$ at $\alpha_{c}$ in the experiment is strongly consistent with that in the theory [Fig. \ref{aJJ} (b)].
Note that the observed excitation above $\alpha_{c}$ was proved to be the triplet [Fig. \ref{FH} (a)], showing the $S_{z}=\pm1$ Zeeman splitting to be similar to those observed below $\alpha_{c}$ .
Although the corresponding excitation was observed at 2 meV (0.48 THz) from INS measurements at 2.15 GPa above $P_{c}$ \cite{INS}, it was obtained at zero field and its property was only discussed from the viewpoint of the structure factor.
In contrast, our observation gives direct evidence that the observed excitation is a triplet.
Thus, the observed transition exhibits the QPT to the plaquette phase.
For the result that the lower excitation D-mode was not observed in the measured range, there is a possibility that the temperature is not sufficiently low or that this mode does not satisfy the ESR selection rule.

Note that the gap remains open at $P_{c}$ in Fig. \ref{FH} (b), which also shows that the observed QPT is first-order \cite{koga, takushima, corboz}, in contrast to the behavior suggested from X-ray measurements \cite{X-ray}.
Haravifarda {\it et al}. evaluated the gap energy by fitting the Arrhenius formula to the temperature dependence of the reciprocal of the lattice parameter, which scales the magnetic susceptibility \cite{X-ray}.
However, the gap energy obtained by this formula is neither relevant \cite{kageyama} nor consistent with our result, which is supported by theoretical results [Fig. \ref{aJJ} (b)].

For the discrepancy between the observed transition point $\alpha_{c}$ and theoretical $\alpha_{c}({\rm theory})$, there are two possible origins.
One is that the model Hamiltonian ${\cal H}$ requires additional terms to describe the real material.
To improve the discrepancy quantitatively, the intradimer DM interaction, the in-plane component of the interdimer DM interaction \cite{DM} or the interplane exchange interaction \cite{miya2, knetter} might be required.
The other is the possibility that the intradimer interaction $J$ of the two orthogonal  dimers become inequivalent by applying the pressure.
The phase diagram for when the intradimer interaction is inequivalent ($J_{1}$, $J_{2}$) in the Hamiltonian (\ref{eq1}) was obtained theoretically and shows that the plaquette phase is robust in some ($J_{2}/J_{1}$, $J'/J_{1}$) regions \cite{moliner}.
In assuming $J_{2}/J_{1}\sim0.95$, the QCP to the plaquette phase was found to be $J'/J_{1}\sim0.66$ from this phase diagram, which corresponds to $\alpha_{c}\sim0.66$ that was obtained in this study.
This explains not only the discrepancy between the observed $\alpha_{c}$ and theoretical $\alpha_{c}({\rm theory})$, but also the loss of four-fold symmetry around the $c$ axis observed in NMR \cite{NMR, NMR2}.

Finally, we discuss the pressure dependence of $J$ and $\alpha$, and the magnetostriction effect on the magnetization plateau in connection with the recently proposed nanopantograph mechanism of this system \cite{radtke}.
From Fig. \ref{aJJ} (a), $\alpha$ increases as pressure increases.
This is because the pressure dependence of $J$ is much larger than that of $J'$, although both decreases as the pressure increases.
This fact is qualitatively consistent with the calculation using density functional theory, from which $J$ decreases as the lattice parameter $a$ decreases, whereas the change in $J'$ is rather small \cite{radtke}.
The nanopantograph mechanism, which originates from the unique orthogonal dimer arrangement of SrCu$_{2}$(BO$_{3}$)$_{2}$, also supports a large reduction in $J$.
It explains why the intradimer Cu-O-Cu angle $\theta$, which governs the magnitude of $J$, can be reduced significantly even with a slight change $\Delta a$ in lattice constant in the $ab$ plane.
The change in angle $\Delta\theta(P)$ is related to the rate of change of the lattice parameter $\Delta a(P)/a$ as $\Delta\theta(P)\simeq35\Delta a(P)/a$ [rad.] \cite{radtke}.
From the X-ray measurements at 4 K, the lattice parameter $a$ was found to decrease linearly as $\Delta a(P)/a=-1.18\times10^{-3}P[{\rm GPa}]$ \cite{X-ray}.
Although at the transition pressure $P_{c} = 1.85$ GPa the corresponding change is  small ($\Delta a/a=-2.2\times10^{-3}$), we obtained a relatively large angular change by the above-mentioned relation ($\Delta\theta=-0.076$ rad. $=-4.4^{\circ}$);
it corresponds to a change in Cu-O-Cu angle from 97.6$^{\circ}$ \cite{crystal} at ambient pressure to 93.2$^{\circ}$ at $P_{c}$.
Hence, there is no doubt about the reduction in the intradimer exchange interaction $J$ according to the Goodenough-Kanamori rule \cite{kanamori}.
Furthermore, the nanopantograph mechanism may also affect the magnetization plateaus because $\alpha$ varies depending on the magnitude of magnetostriction in the $ab$ plane under the field.
Because the theoretical phase diagram, which shows the magnetization plateaus in the $\alpha-H$ plane, is rather complicated around $\alpha$ at ambient pressure just below the QCP \cite{matsuda}, a change in $\alpha$ should be taken into account at each plateau when the experimentally obtained magnetization curve is compared with theory, as suggested by Radke {\it et al}. \cite{radtke}.
We examine this magnetostriction effect on the magnetization plateaus quantitatively.
The relation $\alpha(P)=0.0322P+0.601$ obtained in this study is connected with the $\Delta a(P)/a$ by eliminating the variable $P$ to give $\alpha=-27.3\Delta a/a+0.601$.
If we simply assume that this relation can be applied for the relation between the magnetostriction in the $ab$ plane $\Delta a(H)/a$ and the ratio $\alpha(H)$, we can estimate the change in $\alpha(H)$ resulting from the change in field.
The change is found to be rather small even at a very high field, contrary to the suggestion of Radtke {\it et al}.
For instance, at around 80 T, where the existence of the 2/5 magnetization plateau is controversial \cite{matsuda, jaime}, $\Delta a/a$ is $-2\times10^{-4}$ at most \cite{radtke, jaime} yielding $\alpha-0.601\sim0.005$.
Thus, the obtained relation between $\alpha$ and $\Delta a/a$ helps in furthering our understanding of the controversial magnetization plateaus such as the 2/5 plateau.

In conclusion, we have performed ESR measurements for various pressures, and we succeeded in obtaining the pressure dependence of excitation energies directly.
Performing an exact diagonalization of an appropriate model for the present study, the excitation energies obtained from calculation and experiment were compared, and the pressure dependence of $J$, $\alpha$ and $D$ were determined uniquely to be $J(P)/k_{\rm B}$[K] $= -5.14$[K/GPa]$P$[GPa]$+69.1$[K], $\alpha(P)=0.0322P[{\rm GPa}]+0.601$, and $D(P)=0.039J'(P)$, respectively.
We found that the reduction of the intradimer interaction $J$ contributes mainly to an increase in $\alpha$.
Moreover, this precise determination of these parameters makes it possible to explore the magnetostriction effect on the magnetization plateau in detail.
From a comparison with the theoretical results in Takushima {\it et al} \cite{takushima}., we revealed that the system undergoes a first-order quantum phase transition to the plaquette phase at $\alpha_{c}=0.660\pm0.003$ ($P_{c}=1.85\pm0.05$ GPa) \cite{INS}.

\begin{acknowledgment}

This research was partially supported by Grants-in-Aid for Scientific Research (C) (No. 16K05416) from Japan Society for the Promotion of Science.
\end{acknowledgment}

\end{document}